\documentclass[preprint]{jpsj2}

\usepackage{cases}

\renewcommand{\t}[1]{\tilde{#1}}
\newcommand{\an}[1]{\left\langle #1 \right\rangle}

\newcommand{\Rcpd}{R_c(D)}

\def\vec#1{\mbox{\boldmath $#1$}}

\newcommand{\ltsim}{\protect\raisebox{-0.5ex}{$\:\stackrel{\textstyle <}
	{\sim}\:$}}
\newcommand{\gtsim}{\protect\raisebox{-0.5ex}{$\:\stackrel{\textstyle >}
	{\sim}\:$}}

\newcommand{\sumj}{\sum_{j \in Y}}
\newcommand{\suml}{\sum_{l \in \t{Y}}}

\title
{Statistical Mechanical Approach to Error Exponents 
of Lossy Data Compression}

\author{Tadaaki \textsc{Hosaka}
        \thanks{E-mail address: hosaka@sp.dis.titech.ac.jp} and 
        Yoshiyuki \textsc{Kabashima}
        \thanks{E-mail address: kaba@dis.titech.ac.jp}
}

\inst{Department of Computational Intelligence and
 Systems Science, Tokyo Institute of Technology, Nagatsuta-cho,
 Midori-ku, Yokohama 226-8502}

\abst{
We present herein a scheme by which to accurately evaluate the error
exponents of a lossy data compression problem, which characterize
average probabilities over a code ensemble of compression 
failure and success above or below a critical compression rate, 
respectively, utilizing the replica method (RM). 
Although the existing method used 
in information theory (IT) is, in practice, limited to 
ensembles of randomly constructed codes, the proposed RM-based approach 
can be applied to a wider class of ensembles. 
This approach reproduces the 
optimal expressions of the error exponents 
achieved by the random code ensembles, 
which are known in IT. 
In addition, the proposed framework is used to show that 
codes composed of non-monotonic perceptrons 
of a specific type can provide the optimal 
exponents in most cases, which is supported by  
numerical experiments.
}

\kword{error exponent, lossy data compression,
replica method, 
random code ensemble, perceptron}

\begin{document}
\maketitle


\section{Introduction}
Recent research activities in the cross-disciplinary field 
that combines information theory (IT) and statistical 
mechanics (SM)
have shown that the typical performance of various codes,
such as error correction 
and compression codes,  
can be characterized as phase transitions between 
several phases representing the success or failure of coding 
when the length of messages $M$ becomes 
infinite~\cite{Nishimori}. 
However, for finite $M$, 
probabilities of coding failure 
in the success phase and 
coding success in the failure 
phase do not vanish, and therefore,
it is interesting to estimate the probabilities
that those events occur.

For a reasonable code ensemble, 
the averages of those probabilities over the ensemble asymptotically
scale 
with respect to $M$ as $\exp [-M \alpha]$.
Here, $\alpha (> 0)$ which characterizes the 
asymptotic behavior, is often termed the {\em error exponent}. 
The evaluation of $\alpha$ is theoretically 
interesting and is of practical importance in 
the sense that the error exponents can be useful
as one criterion in the case of assessing the coding 
performance for finite $M$.

More recently, it has been shown that the 
replica method (RM) developed 
in SM can be used for accurate assessment of such an exponent
for error correcting codes \cite{typical,reliability}. 
Nevertheless, the proposed method relies on specific properties 
of error correcting codes, and
the development 
of such techniques for other codes requires
further investigation. 
Therefore, we herein provide a scheme by which 
to accurately evaluate the error exponents for lossy data 
compression problems of memoryless sources utilizing RM. 
The existing method used in IT has provided 
the optimal expressions of the error
exponents \cite{Marton, Csiszar}.
However, a precise assessment 
by the IT approach is, in practice, possible only for the 
ensembles of randomly constructed codes that exhibit
optimal performance. 
In contrast, our SM-based approach can 
accurately evaluate the coding performance for
a wider class of ensembles.

The paper is organized as follows. 
In the next section, we briefly review the concept of  
lossy data compression and 
the definition of the error exponents. 
In {\S}3, a statistical mechanical approach 
for the assessment of the error exponents is introduced. 
In {\S}4, this approach is applied 
to the random code ensemble (RCE).  
Although the exponents evaluated here characterize
the asymptotic behavior of the average probabilities
over the ensemble, this analysis successfully 
reproduces the known optimal exponents in IT literature 
by selecting the best code in that ensemble. 
We briefly discuss RM-based evaluation 
of the exponents of the best code, which reproduces
a result that is identical to the analysis 
for the average case. In addition to being consistent with 
the existing IT results, a major advantage of the proposed method
is the ability to accurately assess the exponents
for suboptimal ensembles. This is demonstrated for a simple 
lossy compression problem of 
a binary memoryless source in {\S}5. 
For this source, the error exponents 
are evaluated for a suboptimal code ensemble, 
composed of perceptrons, of practical codebook size 
using the developed RM-based approach. 
The validity of the assessment is also 
confirmed numerically. 
The final section is devoted to a summary.

\section{Lossy Data Compression and Error Exponents}
In this section, we present the notation used herein and 
briefly review the concept of lossy data compression
of memoryless sources.
Let us focus on a discrete message consisting of $M$ random variables 
$\vec{y} = (y^1,y^2,\ldots,y^M)~(y^\mu \in Y \!=\! \{0,1, \ldots, J-1 \} )$, 
each component of which is assumed to be independently 
generated from an identical stationary distribution 
$\vec{P} = (P(0), P(1), \ldots, P(J-1) )$. 
Although the arguments below
are for sources of discrete messages, 
the newly developed scheme can be directly extended to the case of continuous memoryless sources, 
in which the error exponents are expressed identically by replacing
summations and distribution functions 
with integrals and density functions, respectively.	

The purpose of lossy data compression is 
to compress $\vec{y}$ into a binary expression 
$\vec{s} = (s_1,s_2,\ldots,s_N)~(s_i \in \{0,1\})$, 
allowing a certain amount of {\em distortion} between the 
original message $\vec{y}$ and the representative 
vector $\t{\vec{y}} 
= (\t{y}^1,\t{y}^2,\ldots,\t{y}^M)~(\t{y}^\mu 
\in \t{Y} \!=\! \{0,1, \ldots, L-1 \})$ 
when $\t{\vec{y}}$ is retrieved from $\vec{s}$.
We deal herein with the distortion of 
single-letter fidelity criterion 
$d$ on $Y \times \t{Y}$, which is defined 
as $d(j,l) \ge 0~(j \in Y, l \in
\t{Y})$ and 
$\mathrm{min}_{l \in \t{Y} } \{ d(j,l) \} = 
0~( \forall j \in Y)$. For example,
the distortions for Boolean messages 
$Y=\t{Y}= \{0,1\}$ are frequently measured 
using the Hamming distance, 
$d(\vec{y},\t{\vec{y}}) 
= \sum_{\mu=1}^M  \left[1 - 
\delta_{y^\mu, \t{y}^\mu} \right]  \ge 0,$
where $\delta_{x, y}$ is $1$ if $x=y$, and $0$ otherwise.

A code $\mathcal{C}$ is specified by a map
$\t{\vec{y}}(\vec{s}; \mathcal{C}):\vec{s} 
\to \t{\vec{y}}$, which is
used in the restoration phase.
This reasonably determines the compression scheme as 
\begin{eqnarray}
\vec{s}(\vec{y}; \mathcal{C}) = 
\underset{\vec{s}}{\mathrm{argmin}}
\{ d(\vec{y}, \t{\vec{y}}(\vec{s}; \mathcal{C}))\},
\label{eq:selecting_s}
\end{eqnarray}
where $\mathrm{argmin}_{\vec{s}} \{ \cdots\}$ represents 
the argument $\vec{s}$ that minimizes $\cdots$. 
When ${\mathcal C}$ is generated from 
a certain code ensemble, typical codes 
satisfy the fidelity 
criterion 
\begin{eqnarray}
\frac{1}{M}
\mathop{\rm min}_{\vec{s}}
\{d(\vec{y},\t{\vec{y}}(\vec{s}; \mathcal{C}) \} 
= \frac{1}{M}\mathop{\rm min}_{\vec{s}}
\left\{ \sum_{\mu=1}^M d(y^\mu, \t{y}^\mu(\vec{s};\mathcal{C}))
\right\} < D, 
\label{eq:fidelity}
\end{eqnarray}
for a given permissible distortion $D$ and typical messages 
$\vec{y}$ with probability $1$
in the limit $M,N \to \infty$ keeping 
the coding rate $R \equiv N/M$ constant, if and only if $R$
is larger than a certain critical rate $\Rcpd$.

However, for finite $M$ and $N$,
any code has a finite probability 
$P_{\rm \tiny F}$ of breaking the 
fidelity (\ref{eq:fidelity})
even for $R>\Rcpd$. 
Similarly, for $R <\Rcpd$, 
eq.(\ref{eq:fidelity}) 
is satisfied with a certain probability $P_{\rm \tiny S}$. 
For reasonable code ensembles, 
the averages of these probabilities
are expected to decay exponentially with respect to 
$M$ when the message length $M$ is sufficiently large. 
Therefore, the two error exponents
$\alpha_A (D,R) = \lim_{M \to \infty} 
-\frac{1}{M}
\ln \left \langle P_{\rm \tiny F} 
\right \rangle_{\cal C}$ for $R >\Rcpd$ 
and $\alpha_B (D,R) = \lim_{M \to \infty} 
-\frac{1}{M} \ln\left \langle  P_{\rm \tiny S} 
\right \rangle_{\cal C}$ for $R <\Rcpd$,
where $\langle \cdots \rangle_\mathcal{C}$ represents the average over
the code ensemble,
can be used to characterize
the potential ability of the ensembles of finite 
message lengths. The development of a framework for evaluating these
exponents utilizing RM is the primary goal of this paper.


\section{Statistical Mechanical Approach to Error Exponents}
\subsection{Free energy as a lower-bound of distortion}
Let us develop an analytical framework 
to assess the error exponents using RM.
For this, we first regard the distortion function 
$d(\vec{y}, \t{\vec{y}} (\vec{s}; \mathcal{C}))$ 
as the Hamiltonian for the dynamical variable $\vec{s}$, 
which also depends on predetermined
variables $\vec{y}$ and $\mathcal{C}$. 
In the compression process, 
the optimal sequence is chosen as eq.
(\ref{eq:selecting_s}). As the original message 
and the code are generated from a stationary distribution
$\vec{P}$ and the code ensemble, respectively, 
the resulting distortion (per bit) 
$\lambda(\vec{y},\mathcal{C}) = \mathrm{min}_{\vec{s}}
\{ M^{-1} d(\vec{y}, \t{\vec{y}}(\vec{s}; \mathcal{C}))\}$ 
is also expected to obey a certain 
distribution $P(\lambda, R)$.

\begin{figure}
\begin{minipage}{.5\linewidth}
	\includegraphics[width=6cm]{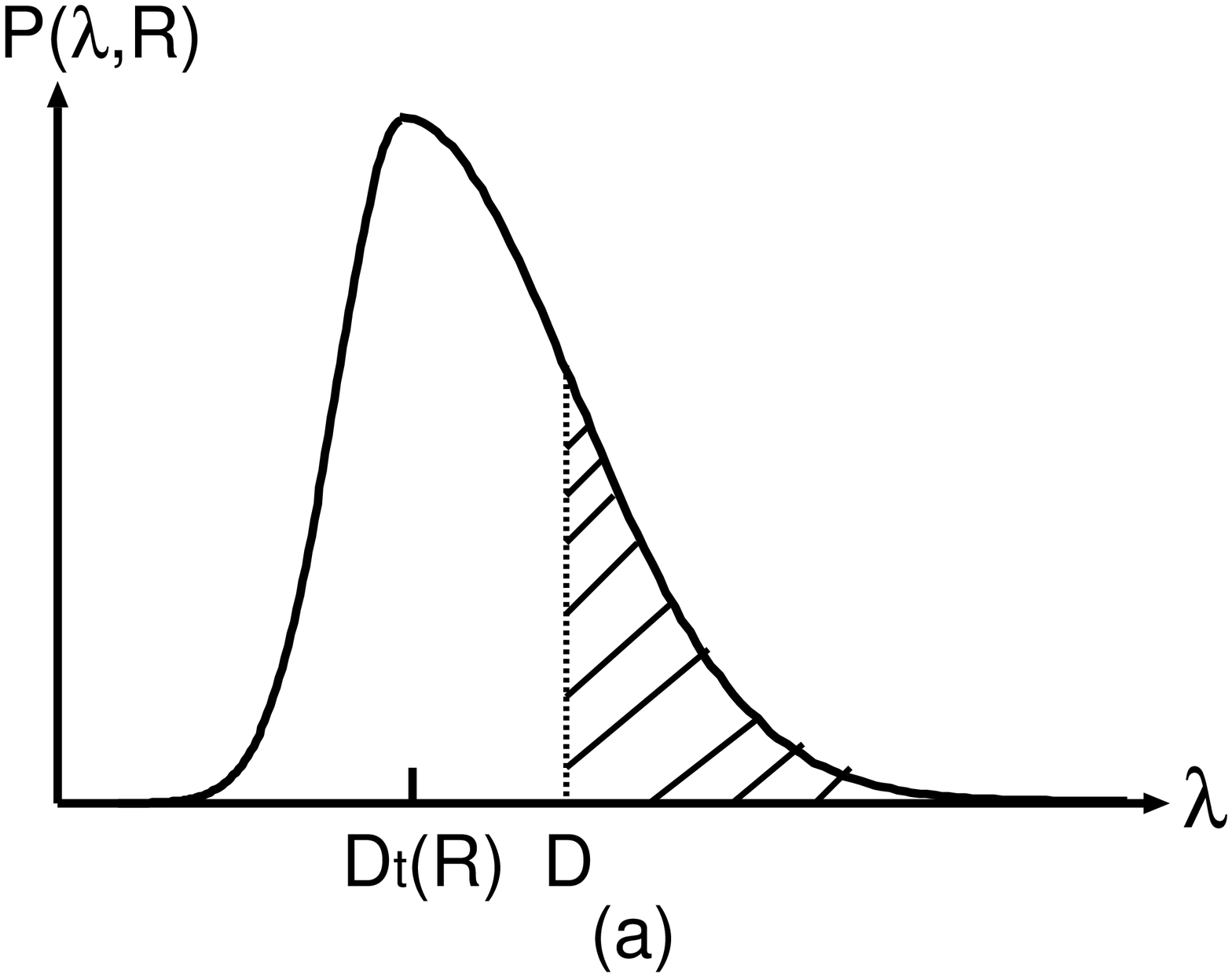}
\end{minipage}
\begin{minipage}{.5\linewidth}
	\includegraphics[width=6cm]{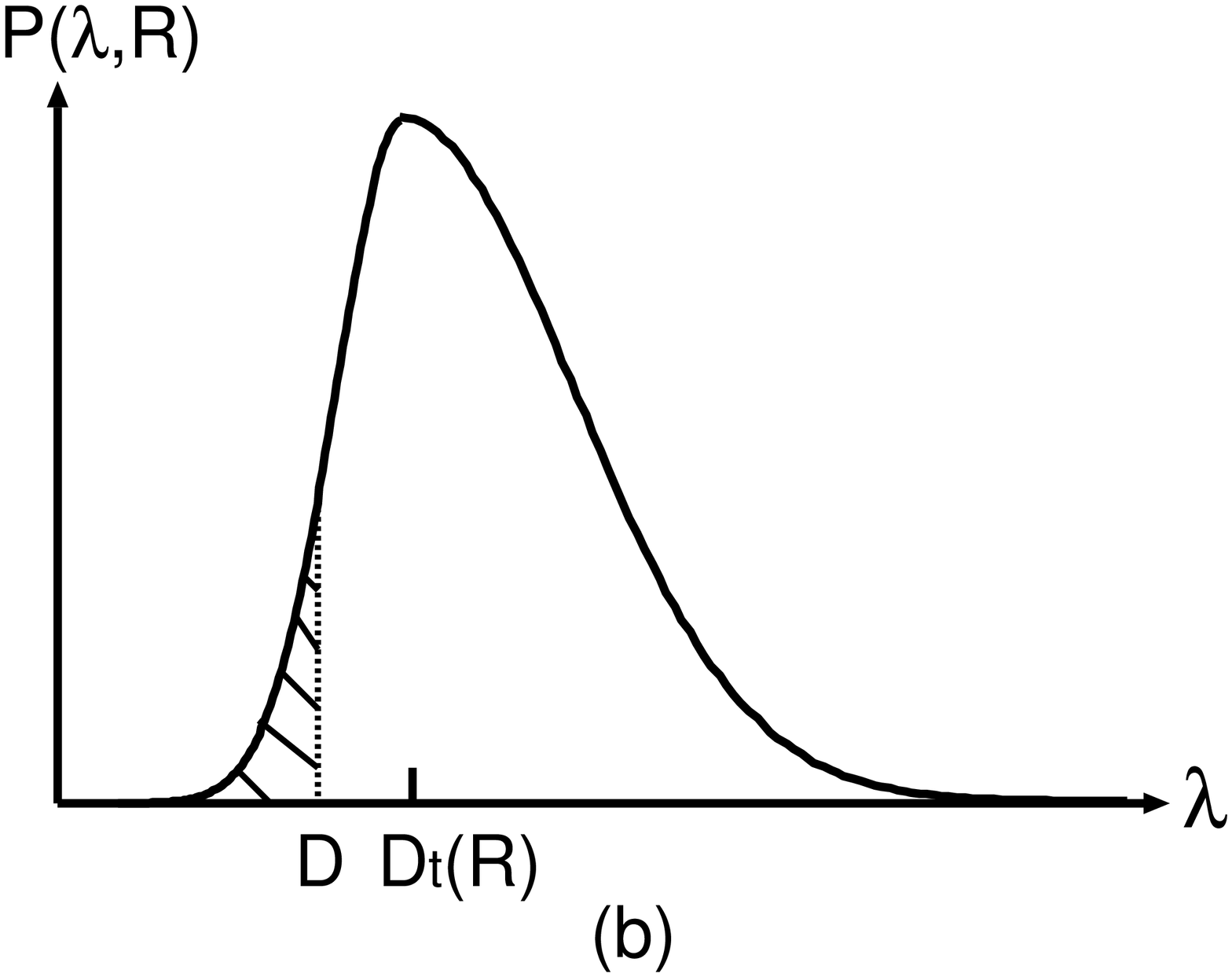}
\end{minipage}
\caption{Schematic profile of 
the distribution $P(\lambda, R)$.
$D_t(R)$ indicates the typical value of the distortion 
$M^{-1}\mathop{\rm min}_{\vec{s}} \{
d(\vec{y},\tilde{\vec{y}}(\vec{s};\mathcal{C})) \}$. 
As $M,N \to \infty$ keeping $R=N/M$ fixed, 
the probability of compression failure $\langle P_{\mathrm{F}} \rangle_{\mathcal{C}}$
for a given $D(> D_t(R))$, which is represented as 
the shadow area (a), tends toward zero. 
Similarly, the probability of compression success 
$\langle P_{\mathrm{S}} \rangle_{\mathcal{C}}$ for a given $D (<D_t(R))$ is illustrated in figure
 (b).
Here, the error exponents are defined for 
characterization of the decay rates of 
these average probabilities. 
}
\label{fig:exponent}
\end{figure}

In the thermodynamic limit, $P(\lambda, R)$ 
is expected to peak at the typical value 
$\lambda=D_t(R)$ and decay exponentially away from 
$D_t(R)$ as $P(\lambda, R) \sim 
\exp[-M h(\lambda, R)]$.
This indicates that $\an{ P_{\mathrm{F}}}_{ \mathcal{C} } = \int_D^\infty
P(\lambda, R) d \lambda \sim 
P(D, R) \sim \exp[-M h(D, R)]$ 
for $D > D_t(R)$ (or $R>\Rcpd$) 
and 
$\an{ P_{\mathrm{S}}}_{ \mathcal{C} } = \int_0^D P( \lambda, R ) d \lambda
\sim P(D, R) \sim \exp[-M h(D, R) ]$ 
for $D < D_t(R)$ (or $R  < \Rcpd$).
Therefore, we can express the error 
exponents $\alpha(D,R)$ using $h(D, R)$
for both cases (Fig. \ref{fig:exponent}).

In order to assess the 
distribution $P(\lambda, R)$, 
we next utilize the inequality
\begin{eqnarray}
e^{-M \beta \lambda(\vec{y}, \mathcal{C})} \le 
\sum_{\vec{s}}e^{-\beta d(\vec{y}, \t{\vec{y}} 
(\vec{s}; \mathcal{C}))}
=Z(\beta;\vec{y},\mathcal{C})=e^{-M\beta f(\beta;\vec{y},\mathcal{C})}, 
\label{ineq}
\end{eqnarray}
which holds for any sets of $\beta \!>\! 0, \vec{y}$ and $\mathcal{C}$. 
The physical implication of this is 
that the ground state energy $\lambda(\vec{y}, \mathcal{C})$ (per component) 
is lower bounded by the free energy $f(\beta; \vec{y}, \mathcal{C})$ (per component)
for an arbitrary temperature $\beta^{-1} >0$. 
In particular, $f(\beta; \vec{y}, \mathcal{C})$ agrees with $\lambda(\vec{y}, \mathcal{C})$ 
in the zero temperature limit $ \beta \to \infty$. 
This means that we can evaluate $P(\lambda, R)$ 
by first assessing the distribution of $f(\beta; \vec{y}, \mathcal{C})$, $P(f; \beta)$, 
for general finite $\beta>0$, 
and then taking the limit $\beta \to \infty$ afterward.
Note that although most of the
quantities appearing in this paper
actually depend on the coding rate $R$, the
dependency is not specified for some quantities such as $P(f; \beta), c(f,
\beta)$, and $g(n, \beta)$.


\subsection{Assessment of the error exponents 
from the moment of the partition function}
\begin{figure}[t]
\begin{center}
\includegraphics[width=6cm]{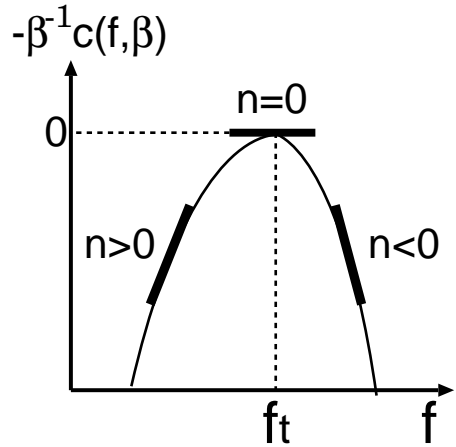}
\caption{Graphical scheme used to solve 
eq.(\ref{eq:saddlecf}), 
enabling the Legendre transformation of eq. 
(\ref{eq:g_Legendre}) to be performed.}
\label{slope}
\end{center}
\end{figure}

$P(f; \beta)$ is also expected to peak at its 
typical value 
\begin{eqnarray}
f_t(\beta) = - \frac{1}{M \beta}
\an{ \ln Z(\beta; \vec{y}, \mathcal{C}) }_{\vec{y}, \mathcal{C}}
= - \lim_{n \to 0} \frac{1}{M \beta} \frac{\partial}{\partial n}
\ln \an{ Z^n(\beta; \vec{y}, \mathcal{C})}_{\vec{y}, \mathcal{C}},
\end{eqnarray}
and decay exponentially away from $f_t(\beta)$ as 
$P(f;\beta) \sim \exp[ - M c(f, \beta)]$ 
for large $M$. 
Here, we assume that $c(f;\beta)\ge 0$ is 
a convex downward function minimized to $0$ at $f=f_t$. 
This implies that, for $\forall{n} \in {\bf R}$, 
the moment of the partition function $Z(\beta;\vec{y},
\mathcal{C})$, 
$\an{Z^n(\beta; \vec{y}, \mathcal{C})}_{\vec{y}, \mathcal{C}}$, 
can be evaluated by the saddle point method as 
\begin{eqnarray}
\an{Z^n(\beta;
\vec{y}, \mathcal{C})}_{\vec{y}, \mathcal{C}}
\approx \exp[ -M \{ n \beta f^* + c(f^*, \beta) \}],
\end{eqnarray}
where $\left \langle \cdots \right \rangle_{\vec{y}, 
\mathcal{C}}$ denotes
the average over $\vec{y}$ and $\mathcal{C}$, and
$f^*$ represents the value at the saddle point, 
which leads to the Legendre transformation
\begin{eqnarray}
g(n, \beta) \equiv -\frac{ \ln 
\an{Z^n (\beta; \vec{y}, 
\mathcal{C} )}_{ \vec{y}, \mathcal{C} }}{M} =
\underset{f}{\mathrm{min} } \{ n \beta f + c(f, \beta) \}. 
\label{eq:g_Legendre}
\end{eqnarray}
Fig. \ref{slope} illustrates graphically the 
meaning of this evaluation.
Given $n$, the minimization 
in eq.(\ref{eq:g_Legendre}) provides
a condition for determining the dominant $f$ as
\begin{eqnarray}
n =-\frac{1}{\beta}\frac{\partial c(f,\beta)}{\partial f}. 
\label{eq:saddlecf}
\end{eqnarray} 
For each $\beta$, this can be 
solved pictorially by searching for the point on $f$
at which the tangential slope of a function 
$y=-\beta^{-1}c(f,\beta)$ agrees with $n$. 
Since $\beta$ is positive, 
$y=-\beta^{-1}c(f,\beta)$ is a convex upward function. 
This indicates that $n=0$, 
$n > 0$ and $n < 0$ correspond to the typical 
values $f=f_t$, $f < f_t$ and $f > f_t$, 
respectively, which provides a useful clue
for assessing the exponents.

Based on eq.(\ref{eq:g_Legendre}), 
the exponent $c(f, \beta)$ that characterizes 
the distribution of free energy 
$P(f;\beta)$ can be assessed by the inverse 
Legendre transformation 
\begin{eqnarray}
c(f, \beta) = \underset{n}{ \mathrm{max} } 
\{ - n \beta f + g(n, \beta) \}, 
\label{eq:Legendre}
\end{eqnarray}
where $\mathop{\rm max}_x \{ \cdots \}$
denotes the maximization of $\cdots$ with respect to $x$, 
from $g(n, \beta)$, which can be evaluated by using RM
analytically extending expressions obtained 
for $n \in {\bf N}$ to $n \in {\bf R}$
if $f$ is included in the support 
of $P(f;\beta)$, which we assume below. 
This enables the evaluation of 
the error exponent $\alpha(D , R)$, 
where $D$ is assumed to be included in the 
support of $P(\lambda,R)$ throughout this paper, 
as $\alpha(D =f, R)=h(\lambda = f, R)=c(f, \beta \to \infty)$ 
taking the zero-temperature limit $\beta \to \infty$. 
The extremum with respect to $n$ 
in eq.(\ref{eq:Legendre}) 
is characterized by the condition 
\begin{eqnarray}
\frac{1}{\beta}\frac{\partial g(n,\beta)}{\partial n}=f,
\label{eq:extremum_n}
\end{eqnarray}
for a given $f$, indicating that 
the exponent $\alpha_{ \{ A,B \} }(D,R)$, 
which is an abbreviation denoting
$\alpha_A(D,R)$ and 
$\alpha_B(D,R)$ for $R >\Rcpd$ and $R
<\Rcpd$,
respectively,
can be assessed as
\begin{eqnarray}
\alpha_{ \{ A,B \} }(D,R) &= &
\lim_{\beta \to \infty} c(f=D, \beta )  \cr
&=& \lim_{\beta \to \infty}
\left \{
-n \frac{\partial g(n, \beta)}{\partial n} + g(n, \beta)
\right \},
\label{eq:exact_exponent}
\end{eqnarray}
where $n$ in eq.(\ref{eq:exact_exponent}) 
is a function of $\beta$ that is determined by the condition 
\begin{eqnarray}
\frac{1}{\beta}\frac{\partial g(n,\beta)}
{\partial n}=D. 
\label{n_condition}
\end{eqnarray}
Equations (\ref{eq:exact_exponent}) and (\ref{n_condition}) 
constitute the basis of our approach. 

It is necessary to mention two points here. 
First, $\alpha_A(D,R)$ is evaluated 
for $R > R_c(D)$, or $D > D_t(R)$
for fixed $R$, where the typical 
distortion $D_t(R)$ can be 
evaluated as $D_t(R)=\lim_{\beta \to \infty} f_t(\beta)$. 
Since Fig. \ref{slope} indicates
that $f > f_t$ corresponds to $ n < 0$, 
$n$ determined from eq.(\ref{n_condition})
becomes negative in the assessment of $\alpha_A(D,R)$. 
Similarly, $n > 0$ is obtained for $\alpha_B(D,R)$. 
Second, we assume that $c(f,\beta)$ is a convex 
downward function of $f$ for $\forall{\beta}$, 
which may not hold in certain situations. 
In such cases, evaluation based on eqs. 
(\ref{eq:exact_exponent}) and (\ref{n_condition}) 
provides the lower bounds of the error exponents 
due to the nature of the Legendre transformation.

\section{Application to the Random Code Ensemble}
\subsection{The random code ensemble}
In order to show that the 
assessment of the error exponents based 
on eqs.(\ref{eq:exact_exponent}) and (\ref{n_condition})
is consistent with the existing results, 
we first apply this method to the random code ensemble (RCE), 
which has been reported extensively 
in IT literature~\cite{Gallager,ViterbiOmura}. 

The RCE is an ensemble that is characterized by the 
component-wise random construction of 
a map $\tilde{\vec{y}}(\vec{s};\mathcal{C})$ 
from $\vec{s}$ to representative sequences 
$\tilde{\vec{y}}$ 
following an identical distribution 
$ \vec{Q} = ( Q(0), Q(1), 
\ldots, Q(L-1) )$, as 
\begin{eqnarray}
{\rm Prob}\left \{ {\tilde{y}}^\mu(\vec{s};\mathcal{C})=
l \right \}
=Q(l), 
\label{randomcodeensemble}
\end{eqnarray}
where $Q(l) \ge 0$ ($l=0,1,\ldots,L-1$)
and $\suml Q( l ) = 1$. 
The correspondence between $\vec{s}$ 
and $\t{y}(\vec{s})$, termed a codebook, is known 
to both the compressor and the decompressor.
The size of the codebook of the RCE
grows as $O(M \times 2^N)$, 
which makes compressing a given message 
computationally difficult when 
the message lengths $N$ and $M$ are large
because, other than looking up the codebook, no compression method exists. This prevents the RCE from being practical. 
However, this ensemble 
exhibits optimal compression 
performance when appropriately tuned, 
and so analysis of the RCE is important for clarifying the 
theoretical limitations of the 
framework of lossy data compression.

\subsection{The replica method: two replica symmetric solutions}
Let us evaluate $g(n,\beta)$ for RCE 
utilizing RM in order to assess 
eqs.(\ref{eq:exact_exponent}) and (\ref{n_condition}). 
For this, we insert an identity $1=\sum_{q_{ab}=1,0} \prod_{a>b}
\delta\left (\delta_{\vec{s}^a,\vec{s}^b} - q_{ab} \right )$
($a,b=1,2,\ldots,n$) into 
$Z^n(\beta; \vec{y}, \mathcal{C})$
for $n \in \mathbf{N}$ and take the averages
over $\vec{y}$ and $\mathcal{C}$, which yields 
\begin{eqnarray}
g(n, \beta)=\mathop{\rm extr}_{ \left\{  q_{ab}\in \{0,1\} \right\} }
\left \{
-\frac{1}{M}
\ln \left [
\mathop{\rm Tr}_{\vec{s}^1,\! \vec{s}^2,\!\ldots, \! \vec{s}^n}
\! \left \langle 
\prod_{a=1}^n  \!
e^{- \beta d(y, \t{y}(\vec{s}^a) )}
\!
\right \rangle_{y,\!\t{y}(\!\vec{s}^1\!),\!\t{y}(\!\vec{s}^2\!), 
\!\ldots,  \t{y}(\!\vec{s}^n\!)}^M \right. \right. \cr
\left. \left . \times \prod_{a>b}
\delta \left (\delta_{\vec{s}^a,\vec{s}^b} - q_{ab} \right )
\right ] \right\}, 
\label{eq:replica}
\end{eqnarray}
where the summation $\sum_{q_{ab}=0,1}$ is replaced with 
the extremization 
$\mathop{\rm extr}_{ \{ q_{ab} \in \{0,1\} \} }$, which is valid 
for $M \to \infty$, 
and $\left \langle \cdots \right \rangle_{y,\!\t{y}(\!\vec{s}^1\!),\!\t{y}(\!\vec{s}^2\!), 
\!\ldots, \t{y}(\!\vec{s}^n\!)}$
denotes the averages over the distributions
$\vec{P}$ and $\{ \vec{Q}( \t{y}(\vec{s}^a )) \}$
$(a=1,\!2,\!\ldots,\!n)$.

In order to utilize this expression for real 
(and, more generally, complex) 
$n$, we first employ the simplest 
replica symmetric (RS) ansatz 
$q_{ab}=q$ $(a > b \! = \! 1, \! 2,\ldots,n)$. 
The value of $q$ is limited to 
only $0$ or $1$ in the current 
system, yielding two RS solutions:
\begin{eqnarray}
g_{\rm \tiny RS1}(n, \beta) 
= 
- \ln \left[ 
\sumj P(j) \left\{
\suml Q(l) e^{ -\beta d(j,l)}
\right\}^n
\right]
- n R \ln 2,
\label{eq:sol0}
\end{eqnarray}
and 
\begin{eqnarray}
g_{\rm \tiny RS2}(n, \beta)  
= 
- \ln \left[ 
\sumj P(j) 
\suml Q(l) 
e^{ -n \beta d(j,l)}
\right]
- R \ln 2,
\label{eq:sol1}
\end{eqnarray}
which correspond to $q\!=\!0$ and $1$, respectively.

\subsection{Critical conditions and the 
frozen replica symmetry breaking solution}
We now have two RS solutions: (\ref{eq:sol0})
and (\ref{eq:sol1}). These solutions, however, become invalid 
unless both of the following two conditions 
are satisfied, which signals the breakdown of the RS ansatz. 

The first condition is regarding the local stability of 
the RS saddle point with respect to the infinitesimal 
disturbance for breaking the replica symmetry
in order parameters, which is often termed the 
de Almeida-Thouless (AT) condition \cite{AT}. 
However, such a disturbance is not allowed 
in the current system, since the order parameters
$q_{ab}=\delta_{\vec{s}_a,\vec{s}_b}$ 
are discrete. Therefore, 
we expect that the AT stability is always satisfied 
for both of the solutions for the RCE, although 
the stability must be examined for other code ensembles. 

The other condition is regarding the entropy of 
the dynamical variable $\vec{s}$. 
Equations (\ref{ineq}) and 
(\ref{eq:g_Legendre}) indicate that 
the equality
\begin{eqnarray}
s(n,\beta)&=&
-\frac{\partial g(n, \beta)}{\partial n} 
+ \frac{\beta}{n} \frac{\partial g
(n, \beta)}{\partial \beta} \cr
&=&
\frac{1}{M} 
    \frac{ 
         \an{ Z^n (\beta; \vec{y}, \mathcal{C}) 
         \left\{  \ln Z(\beta; \vec{y}, \mathcal{C}) 
                - \beta \frac{\partial \ln Z(\beta; \vec{y}, \mathcal{C}) }
          { \partial \beta}
         \right\}
          }_{\vec{y}, \mathcal{C}}
         }
         {   \an{ Z^n (\beta; \vec{y}, \mathcal{C}) }_{\vec{y}, \mathcal{C}}
         }
\label{entropy}
\end{eqnarray}
holds for any pairs of $n$ and $\beta > 0$. 
Since $\ln Z(\beta; \vec{y}, \mathcal{C}) 
- \beta \frac{\partial \ln Z(\beta; \vec{y}, \mathcal{C})}
{\partial \beta} $
represents the entropy of the discrete dynamical variable 
$\vec{s}$ given $\vec{y}$ and $\mathcal{C}$, 
eq.(\ref{entropy}) must become non-negative 
as long as $g(n,\beta)$ is correctly evaluated. 

Substituting eq.(\ref{eq:sol1}) into 
eq.(\ref{entropy}) yields $s(n,\beta)=0$, which 
indicates that $g_{\rm \tiny RS2}(n,\beta)$ always 
critically satisfies this entropy condition. 
However, for $g_{\rm \tiny RS1}(n, \beta)$, 
$-\frac{\partial g_{\rm \tiny RS1}(n, \beta)}{\partial n} 
+ \frac{\beta}{n} \frac{\partial g_{\rm \tiny RS1}(n, \beta)}
{\partial \beta} $ generally vanishes at a certain critical value 
$\beta=\beta_c$, signaling the breakdown 
of the replica symmetry when $\beta$ is increased.
In such cases, one promising method 
for obtaining the correct solution 
is to employ the \textit{1-step (frozen) 
replica symmetry breaking (1RSB) ansatz}, 
partitioning the replicated systems into 
$\frac{n}{m}$ subgroups of identical size $m$ 
and assuming that $q_{ab}=1$ if $a$ and $b$ belong 
to the same subgroup, and $0$ otherwise \cite{1RSB}. 
Extremizing $\left \langle Z^n(\beta;\vec{y},\mathcal{C}) \right 
\rangle_{\vec{y},\mathcal{C}}$ with respect to $m$
yields the 1RSB solution for $\beta > \beta_c$ as
\begin{eqnarray}
g_{\rm \tiny 1RSB}(n,\beta)=
\mathop{\rm extr}_{m} \left \{g_{\rm \tiny RS1}\left(\frac{n}{m},m \beta \right)
\right \}=g_{\rm \tiny RS1}(n^*,\beta^*)
\label{eq:1rsb}
\end{eqnarray}
where $n^*$ and $\beta^*$ are assessed from the coupled equations 
\begin{eqnarray}
&&n^* \beta^*=n \beta, \label{rsbcond1} \\
&&-\frac{\partial g_{\rm \tiny RS1}(n^*,\beta^*)}{\partial n^*}
+\frac{\beta^*}{n^*} \frac{\partial g_{\rm \tiny RS1}(n^*,\beta^*)}
{\partial \beta^*} =0, 
\label{rsbcond2}
\end{eqnarray}
which guarantees that $s(n,\beta)$ is non-negative (zero)
for $g_{\rm \tiny 1RSB}(n,\beta)$. 

Equations (\ref{rsbcond1}) and (\ref{rsbcond2}) 
indicate that eq.(\ref{n_condition}) for 
$g_{\rm \tiny 1RSB}(n,\beta)$ is reduced to 
a condition of $g_{\rm \tiny RS1}(n,\beta)$ as
\begin{eqnarray}
\frac{1}{\beta}\frac{\partial g_{\rm \tiny 1RSB}(n,\beta)}
{\partial n} =\frac{1}{\beta^*}\frac{\partial g_{\rm \tiny RS1}
(n^*,\beta^*)}
{\partial n^*}=D. 
\label{reduction_to_rs1}
\end{eqnarray}
Equations (\ref{eq:1rsb}), (\ref{rsbcond2}) and 
(\ref{reduction_to_rs1}) indicate that 
the error exponents can be practically evaluated 
{\em without using the 1RSB solution} as
\begin{eqnarray}
\alpha_{ \{ A,B \} }(D,R) &= &
\lim_{\beta \to \infty}
\left \{
-n \frac{\partial g_{\rm \tiny 1RSB}(n, \beta)}{\partial n} + 
g_{\rm \tiny 1RSB}(n, \beta)
\right \} \cr
&=&-n \frac{\partial g_{\rm \tiny RS1}(n, \beta)}{\partial n} + 
g_{\rm \tiny RS1}(n, \beta), 
\label{RSexponents}
\end{eqnarray}
where $n$ and $\beta$ are determined by
\begin{eqnarray}
&&\frac{1}{\beta} \frac{ \partial g_{\rm \tiny RS1}(n,\beta)}{
\partial n}=D, \label{rsD}\\
&&-\frac{\partial g_{\rm \tiny RS1}(n,\beta)}{\partial n}
+\frac{\beta}{n} \frac{\partial g_{\rm \tiny RS1}(n,\beta)}
{\partial \beta} =0,
\label{rsS}
\end{eqnarray}
when $g_{\rm \tiny RS1}(n,\beta)$ is selected as 
the relevant solution, 
despite the fact that $g_{\rm \tiny RS1}(n,\beta)$ becomes 
invalid for $\beta \to \infty$.

\subsection{Assessment of the error exponents}
We are now ready to evaluate $\alpha_{\{A,B\}}(D,R)$ for the RCE
using the two RS solutions. We first consider 
the failure exponent $\alpha_{A}(D,R)$ for $R > \Rcpd$. 
\subsubsection{$\alpha_{A}(D,R)$}

In order to assess this exponent, we must select 
the relevant solution from $g_{\rm \tiny RS1}(n,\beta)$
and $g_{\rm \tiny RS2}(n,\beta)$. 
Note that 
$g_{\rm \tiny RS2}(n,\beta)$ must not be relevant 
for $n \le 0$ because this solution does not 
satisfy the trivial identity 
\begin{eqnarray}
\lim_{n \to 0} g(n,\beta)=-\lim_{n \to 0} \frac{1}{M} \ln 
\left \langle Z^n(\beta;\vec{y},\mathcal{C}) \right \rangle
=-\frac{1}{M}\ln \left \langle 1 \right \rangle
=0, 
\label{n0limit}
\end{eqnarray}
and therefore the analytic continuation of this solution 
from $n \in \mathbf{N}$ to $n < 0$ is not reliable.
$\alpha_{A}(D,R)$ corresponds to $n \le 0$, and therefore 
we adopt $g_{\rm \tiny RS1}(n,\beta)$ for 
the evaluation of $\alpha_{A}(D,R)$.

Inserting (\ref{eq:sol0}) into 
eqs.(\ref{rsD}) and (\ref{rsS}) yields
\begin{eqnarray}
\sumj U_1(j)  \ln \left[
\suml Q(l) e^{-\beta d(j,l) }\right]
+ R \ln2 + \beta D = 0,     
\label{eq:dgdn1}    \\
\sumj U_1(j)
\suml V_1(l | j) d(j,l) = D,
\label{eq:dgdb1}
\end{eqnarray}
where
the probability distributions 
$\vec{U}_1=(U_1(0), U_1(1), \ldots,
U_1(J-1) )$ and 
$\vec{V}_1= \{ V_1(l | j) \}~(j \in Y, l \in \t{Y})$ 
are defined as
\begin{eqnarray}
U_1(j) &=& 
\frac{
P(j) \left\{  \suml
Q(l)  e^{-\beta d(j,l) }  \right\}^n 
}
{
\sumj
P(j) \left\{  \suml
Q(l)  e^{-\beta d(j,l) }  \right\}^n
},
\label{eq:U1}  \\
V_1(l|j) &=& 
\frac{
Q(l) e^{-\beta d(j,l)}
}
{
\suml
Q(l) e^{-\beta d(j,l)}
}.
\label{eq:V1}
\end{eqnarray}
Inserting eqs.(\ref{rsD}), (\ref{eq:sol0}) 
and (\ref{eq:dgdn1}) into eq.(\ref{RSexponents}) 
yields the following expression for the error exponent
\begin{eqnarray}
\alpha_A (D, R) 
= 
\sumj U_1(j) \ln \left[
\frac{U_1(j)}{P(j)}
\right]
\equiv KL(\vec{U}_1 || \vec{P}),
\label{eq:KL_exponent}
\end{eqnarray}
where $KL(\cdot||\cdot)$ is termed the Kullback-Leibler 
divergence \cite{Cover}.

Equation (\ref{eq:KL_exponent}) characterizes 
the average performance of the RCE specified by $\vec{Q}$. 
Therefore, the performance can be improved by 
maximizing eq.(\ref{eq:KL_exponent}) 
with respect to $\vec{Q}$ under the 
constraint $\sum_{l}Q(l)=1$ and $Q(l) \ge 0$, which reduces to 
\begin{eqnarray}
\sumj
U_1(j)  \frac{e^{-\beta d(j,l)}}
{
\suml 
Q(l) e^{-\beta d(j,l)}
}
=1   \Longrightarrow
Q(l) = \sumj U_1(j) V_1(l|j),~~~~~~\forall l \in \t{Y}.
\label{eq:Q1}
\end{eqnarray}

The set of $n$, $\beta$ and $\vec{Q}$ 
that optimizes the exponent given $D$ and $R$ 
can be searched by the following scheme, 
which is often termed the Arimoto-Blahut algorithm 
(ABA) \cite{Arimoto,Blahut,Cover}. 
We begin with initial conditions 
of $n(<0)$, $\beta(>0)$ and $\vec{Q}$. 
Keeping $\vec{Q}$ fixed, $n$ and $\beta$ 
are first updated by solving eqs.(\ref{eq:dgdn1}) and (\ref{eq:dgdb1})
with respect to these variables, 
which yields $\vec{U}_1$ and $\vec{V}_1$ 
using eqs.(\ref{eq:U1}) and (\ref{eq:V1}). 
Next, $\vec{Q}$ is updated from 
eq.(\ref{eq:Q1}) using the 
obtained $\vec{U}_1$ and $\vec{V}_1$. 
These procedures are iterated until 
$n$, $\beta$ and $\vec{Q}$ converge, 
which is guaranteed by the convexity of 
the mutual information \cite{Cover}. 
Then, the optimized exponent for the given 
$D$ and $R$ is obtained by substituting 
the convergent solution into eq.(\ref{eq:KL_exponent}).

In practice, it is much more convenient to 
deal with $n$ and $\beta$ as control parameters, 
rather than $D$ and $R$, for which $D$ and $R$ are 
easily obtained from eqs.(\ref{eq:dgdn1}) 
and (\ref{eq:dgdb1}), 
after solving $\vec{Q}$ for the given $n$ and $\beta$
by simply iterating eqs.(\ref{eq:U1}), (\ref{eq:V1}) 
and (\ref{eq:Q1}). Inserting the optimal $\vec{U}_1$, 
which is given by the solved $\vec{Q}$ via 
eq.(\ref{eq:U1}), into eq.(\ref{eq:KL_exponent}) 
and varying $n$ and $\beta$, the $\alpha_A(D,R)$ surface is swept out.

\subsubsection{$\alpha_B(D,R)$} 
Next, we turn to the success exponent $\alpha_B(D,R)$
for $R<\Rcpd$.
Since we expect that $g(n,\beta)$ is analytic,
except for a few possible singular points of $n$, 
$g_{\rm \tiny RS1}(n,\beta)$ is likely to be 
relevant for $n \ge 0$ as well, 
at least in the vicinity of $n=0$, 
because this solution is supposed to 
be relevant for $n \le 0$. 
Then, the exponent is obtained as 
$\alpha_B(D,R)=KL(\vec{U}_1 || \vec{P})$, 
which is similar to $\alpha_A(D,R)$. 

\begin{figure}
\begin{minipage}{.47\linewidth}
	\includegraphics[width=6cm]{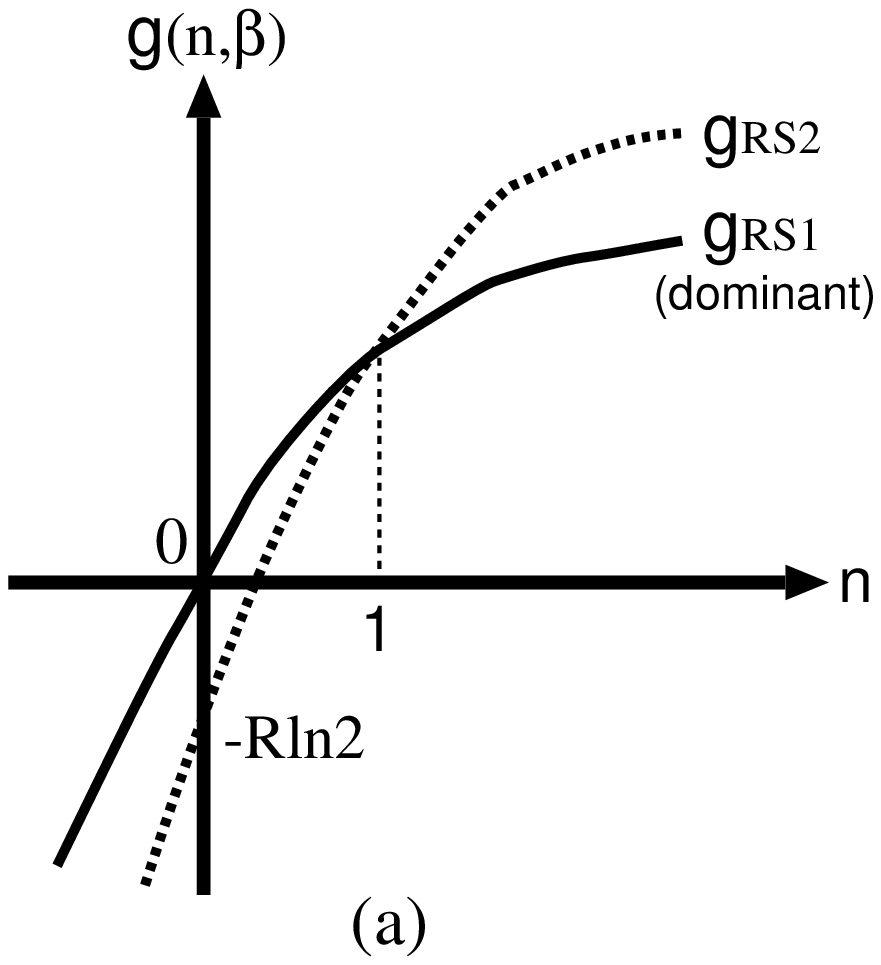}
\end{minipage}
\hspace{0.3cm}
\begin{minipage}{.47\linewidth}
	\includegraphics[width=6cm]{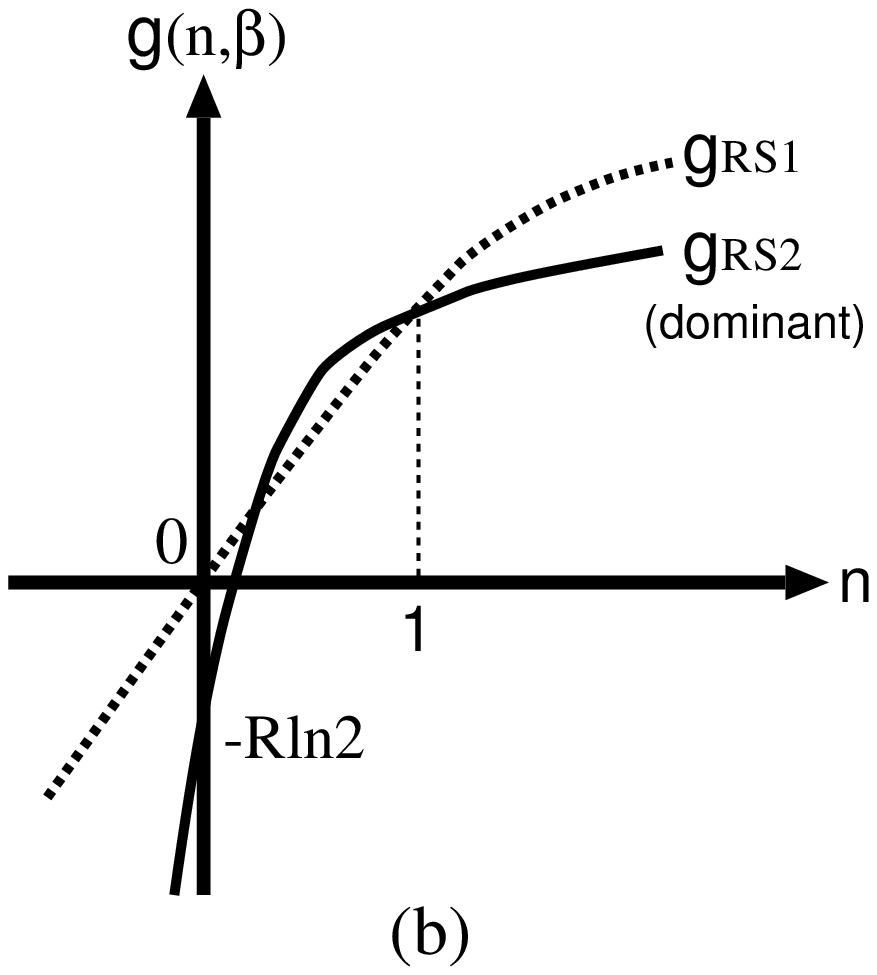}
\end{minipage}
\caption{Schematic profiles of $g_{\mathrm{RS}1}(n,\beta)$ 
and $g_{\mathrm{RS}2}(n,\beta)$ for a fixed $\beta$.
The two functions intersect at $n=1$ and 
both functions are convex upward with respect to $n$.
Whereas the first derivative of $g_{\mathrm{RS}2}(n,\beta)$ 
is always positive, that of $g_{\mathrm{RS}1}$ 
depends on $R, D$ and $n$. RM assesses the value 
of $g(n,\beta)$ for $n \notin \mathrm{N}$ by analytically
continuing the evaluation for $n \in \mathrm{N}$. 
This implies that the relevant solution 
for $n < 1$ is the smaller slope at 
$n=1$ between $g_{\mathrm{RS}1}(n,\beta)$ 
and $g_{\mathrm{RS}2}(n,\beta)$ 
unless the analyticity is broken. 
Thus, the relevant solution is 
$g_{\mathrm{RS}1}(n,\beta)$ and 
$g_{\mathrm{RS}2}(n,\beta)$ for the 
cases of (a) and (b), respectively. 
}
\label{fig:profile_g}
\end{figure}

However, the validity 
of this expression in the present case must be examined  
because $g_{\rm \tiny RS2}(n,\beta)$ 
can be relevant for $n > 0$. 
For this, we illustrate schematic profiles of 
$g_{\rm \tiny RS1}(n,\beta)$ and 
$g_{\rm \tiny RS2}(n,\beta)$ 
for a fixed $\beta$ in Fig. \ref{fig:profile_g}. 

Equations (\ref{eq:sol0}) and (\ref{eq:sol1})
indicate that both $g_{\rm \tiny RS1}(n,\beta)$
and $g_{\rm \tiny RS2}(n,\beta)$, 
which intersect each other at $n=1$ for $\forall{\beta}>0$, 
are convex upward 
with respect to $n$. 
As a function of $n$, $g_{\rm \tiny RS2}(n,\beta)$
increases monotonically. Although the first 
derivative of $g_{\rm \tiny RS1}(n,\beta)$ 
can be both positive and negative, in accordance with eq.(\ref{rsD}), 
only the region of positive slope need to be considered. 

For $n \in \mathbf{N}$, the relevant solution 
of $g(n,\beta)$ can be chosen by selecting 
one of the lower values of the 
two RS solutions, following the criterion 
of the conventional saddle point method. 
For $n \notin \mathbf{N}$, RM relies on the assumption 
that an analytical expression of $g(n,\beta)$ 
that is relevant for a certain natural number 
$k$ is also relevant in the vicinity of $k$,
unless the analyticity is lost~\cite{Ogure}. 
Since $g_{\rm \tiny RS1}(n,\beta)=g_{\rm \tiny RS2}(n,\beta)$
holds at $n=1$, this implies that 
the selection of $g_{\rm \tiny RS1}(n,\beta)$ 
for $n \gtsim 0$, which we tentatively adopted 
assuming that the analyticity of $g(n,\beta)$ is 
not broken between $n < 0$ and $n \gtsim 0$, 
is valid if 
\begin{eqnarray}
&& \left .  
\frac{\partial g_{ \mathrm{RS}2 }(n,\beta) }
{ \partial n} \right|_{n=1}   
- 
\left . 
\frac{\partial g_{ \mathrm{RS}1 } (n,\beta)}
{ \partial n} \right|_{n=1} \cr
&=&
-\frac{1}{M} \frac{\sumj P(j) \suml Q(l) e^{-\beta d(j,l) } (-\beta d(j,l) )}
{\sumj P(j) \suml Q(l) e^{-\beta d(j,l)}} 
\cr
& & 
+ \frac{1}{M} \frac{\sumj P(j) \suml Q(l) e^{-\beta d(j,l) } 
                          \ln\{ \suml Q(l) e^{-\beta d(j,l) }\} }
{\sumj P(j) \suml Q(l) e^{-\beta d(j,l)}}  + R \ln 2 \cr
&=& \left . 
\left (-\frac{\partial g_{\rm \tiny RS1}(n, \beta)}{\partial n} 
+ \frac{\beta}{n} \frac{\partial g_{\rm \tiny RS1}
(n, \beta)}{\partial \beta} \right ) 
\right |_{n=1}> 0, 
\label{validityRS1}
\end{eqnarray}
holds, which corresponds to the situation illustrated
in Fig. \ref{fig:profile_g} (a). 

Let us denote the solution of eqs.(\ref{rsD}) and (\ref{rsS})
as $n=n_c$ and $\beta=\beta_c$, respectively. 
As $\left . \left (-\frac{\partial g_{\rm \tiny RS1}
(n, \beta)}{\partial n} 
+ \frac{\beta}{n} \frac{\partial g_{\rm \tiny RS1}
(n, \beta)}{\partial \beta} \right )
\right |_{n=n_c,\beta=\beta_c}=0$ holds, 
eq.(\ref{validityRS1}) validates the selection of 
$g_{\rm \tiny RS1}(n, \beta)$, which yields
the expression $\alpha_B(D,R)= KL(\vec{U}_1 || \vec{P})$ 
if $n_c < 1$ is obtained, 
because $-\frac{\partial g_{\rm \tiny RS1}(n, \beta)}{\partial n} 
+ \frac{\beta}{n} \frac{\partial g_{\rm \tiny RS1}
(n, \beta)}{\partial \beta} $ is supposed 
to be positive for $n>n_c$ under the 
RS ansatz, which implies that eq.(\ref{validityRS1}) holds. 
However, if $n_c > 1$, eq.(\ref{validityRS1})
does not hold, indicating the situation 
illustrated in Fig. \ref{fig:profile_g} (b). 
In such a case, $g_{\rm \tiny RS1}(n,\beta)$ is no longer relevant for 
$n=n_c$ and $\beta=\beta_c$, 
and therefore we have to amend the solution 
using $g_{\rm \tiny RS2} (n,\beta)$.

For $g_{\rm \tiny RS2} (n,\beta)$, 
eq.(\ref{n_condition}) is given as 
\begin{eqnarray}
\sumj \suml 
V_2(l | j)  U_2(j)
d(j,l) = D,     
\label{eq:dgdn2}
\end{eqnarray}
where distributions $\vec{U}_2=(U_2(0), U_2(1), \ldots,
U_2(J-1) )$ and 
$\vec{V}_2= \{ V_2(l | j) \}~(j \in Y, l \in \t{Y})$ 
are defined as 
\begin{eqnarray}
U_2(j)  
&=& 
\frac{
P(j) \suml Q(l) e^{-\beta^{'} d(j,l) } 
}
{ 
\sumj \suml 
P(j) Q(l) e^{- \beta^{'} d(j,l) }
}, \label{eq:U2} \\
V_2(l|j) 
&=&
\frac{
Q(l) e^{-\beta^{'} d(j,l) } 
}
{ 
\suml 
Q(l) e^{-\beta^{'} d(j,l) }
}, \label{eq:V2}
\end{eqnarray}
respectively, where $\beta^\prime \equiv n \beta$. 

Note that the value of $\beta^\prime $ 
determined from eq.(\ref{eq:dgdn2})
is kept invariant when $\beta$ tends toward infinity. 
In order to assess eq.(\ref{eq:exact_exponent})
for $g_{\rm \tiny RS2} (n,\beta)$,
inserting eq.(\ref{eq:dgdn2})
yields the expression
\begin{eqnarray}
\alpha_B (D,R) = KL( \vec{U}_2 || \vec{P} ) + I - R \ln 2,
\label{exponentRS2}
\end{eqnarray}
where $I$ is defined as
\begin{eqnarray}
I
=
\sumj \suml 
V_2(l|j)  U_2(j)
\ln \left[
\frac{
V_2(l|j) 
}
{
Q(l)
}
\right]. 
\label{eq:I}
\end{eqnarray}
In summary, the exponent 
$\alpha_B (D,R)$ for $R < \Rcpd$ 
is expressed as 
\begin{eqnarray}
\alpha_B (D,R) =
\left \{
\begin{array}{ll}
KL( \vec{U}_1 || \vec{P} ), & \mbox{if $0 <n_c < 1$} \\
KL( \vec{U}_2 || \vec{P} ) + I - R \ln 2, 
&\mbox{if $n_c > 1$} 
\end{array}
\right .
\label{alphaB} 
\end{eqnarray}
for a given ensemble specified by $\vec{Q}$. 

Here, $\alpha_B(D,R)$ can be minimized with respect to
the distribution $\vec{Q}$ 
in a manner similar to that for $\alpha_A(D,R)$. 
Namely, we tentatively adopt the first expression 
of eq.(\ref{alphaB}), assuming that $g_{\rm \tiny RS1}(n,\beta)$
is relevant, and employ ABA 
in order to obtain the optimal $n$, $\beta$ and $\vec{Q}$. 
If the obtained solution of $n$, $n_c$, is smaller than $1$, 
then the obtained expression is appropriate. Otherwise, 
we have to amend the solution 
using the second expression, which can be 
optimized by ABA as well. 
In this case, the convergent solution
satisfies the relation $Q(l)=\sumj V_2(l|j)U_2(j)$
for $\forall{l}\in \tilde{Y}$. This yields the 
expression of the optimized exponent as 
\begin{eqnarray}
\alpha_B (D,R) = KL( \vec{U}_2 || \vec{P} ) + 
(R(\vec{U}_2, D) - R) \ln 2, 
\end{eqnarray}
where 
\begin{eqnarray}
R(\vec{U}, D) = 
\underset{
\underset{\sum_{j\in Y, l \in \tilde{Y}} V(l|j) U(j) d(j,l) 
\le D}{\vec{V}:}}
{ \mathrm{min}}
\sum_{j\in Y, l \in \tilde{Y}} 
V(l|j)  U(j)
\log_2 \left[
\frac{
V(l|j) 
}
{
\sumj V(l|j)U(j)
}
\right],
\label{RDF}
\end{eqnarray}
is termed the {\em rate-distortion function}, 
which represents the theoretically achievable limit 
of the compression rate for the information source 
$\vec{U}$ when distortion up to $D$ is allowed 
in the limit $N,M \to \infty$~\cite{RD}.

\subsection{Consistency with the IT literature}
We obtained two expressions for the error exponents 
(\ref{eq:KL_exponent}) and (\ref{alphaB}) using RM. 
In order to validate our results, 
we check for consistency with 
results in the IT literature. 
\subsubsection{$\alpha_A(D,R)$}

We first examine $\alpha_A(D,R)$ for 
$R > \Rcpd$. In the IT literature, 
the exponent for the best code is provided \cite{Marton} as 
\begin{eqnarray}
\alpha_A^* (D,R) = \underset{\vec{U} : R \le R(\vec{U},D)}
{ \mathrm{min}}     KL(\vec{U} ||\vec{P}).  \label{eq:Marton}
\end{eqnarray}

This minimization problem can be solved by 
the method of Lagrange multipliers.
Introducing auxiliary variables $z_1(\ge 0)$ and 
$z_2(\le 0)$ as
\begin{eqnarray}
z_1 & \equiv & \sumj \suml V(l|j) U(j) \ln \left[
\frac{
V(l|j) 
}
{
Q(l)
}
\right]
- R \ln 2, \label{eq:z1} \\
z_2 & \equiv & \sumj \suml V(l|j) U(j) d(j,l) - D, \label{eq:z2}
\end{eqnarray}
where
\begin{eqnarray}
Q(l)=\sumj V(l|j)U(j),
\label{eq:Inf_Q}
\end{eqnarray}
eq.(\ref{eq:Marton}) is converted to 
the minimization problem of 
\begin{eqnarray}
J_A(\vec{U}, \vec{V},z_1,z_2) &=&
\sumj U(j) \ln \left[ \frac{U(j)}{P(j)} \right] \cr
& & + n_A \left[ \left\{
\sumj \suml V(l|j) U(j) \ln \left[ \frac{V(l|j)}{Q(l)}
\right]  -R \ln 2 - z_1 \right\} \right. \cr
& & + \beta_A \left\{
\sumj \suml V(l|j) U(j) d(j,l) - D - z_2
\right\}  \cr
& &+ \left. \sumj  \nu (j) \left\{
\suml V(l|j) - 1
\right\}
\right]
+ \xi \left\{ \sumj U(j) - 1 \right\},  \label{eq:J}
\end{eqnarray}
with respect to $\vec{U}, \vec{V}, 
z_1$, and $z_2$, where $n_A, \beta_A,
\nu(j)~(j=1,2,\ldots,J-1)$, and $\xi$ are Lagrange multipliers. 

Note that if the minimum is achieved by an
internal point $z_1 >0$, $\partial J_A/ \partial z_1 =
-n_A = 0$; otherwise, the minimum is placed on the boundary
$z_1 =0$ and $\partial J_A / \partial z_1 = -n_A \ge 0$.
Since the rate-distortion function $R(\vec{U}, D)$ decreases monotonically 
as $D$ increases \cite{Cover} and $R >\Rcpd$ is assumed, 
we cannot set $\vec{U} = \vec{P}$.
Furthermore, taking the convexity of the $KL$ 
divergence into account, minimization
(\ref{eq:Marton}) must be achieved on the boundary, 
which ensures that $n_A < 0~(z_1 \!=\! 0)$.
A similar argument holds for $\beta_A$.
According to the convexity of the mutual information,
the rate-distortion function $R(\vec{U},D)$ is determined by the
distribution $\vec{V}$ on its boundary, 
which indicates $\beta_A >
0~(z_2=0)$.

Minimizing eq.(\ref{eq:J}) 
with respect to $V(l|j)$ provides
\begin{eqnarray}
V(l|j) = \frac{ Q(l) e^{- \beta_A d(j,l)}}{ \suml Q(l) e^{- \beta_A d(j,l)}
},~~~~~\forall j \in Y, l \in \t{Y}, \label{eq:Inf_V}
\end{eqnarray}
where the normalization constraint has been already factored into the equation.
Similarly, minimization with respect to
$U(j)$ yields
\begin{eqnarray}
U(j) = 
\frac{
P(j) \left\{  \suml
Q(l)  e^{-\beta_A d(j,l) }  \right\}^{n_A}
}
{
\sumj
P(j) \left\{  \suml
Q(l)  e^{-\beta_A d(j,l) }  \right\}^{n_A}
},~~~~~\forall j \in Y.
\label{eq:Inf_U}
\end{eqnarray}

In practice, we can assess the optimal exponents 
using ABA to
solve eqs.(\ref{eq:Inf_U}),(\ref{eq:Inf_V}) and 
(\ref{eq:Inf_Q}) with respect to $n_A <0$, 
$\beta_A >0$ and $\vec{Q}$ under the constraint 
that the solutions should be found on the boundary, 
which is represented as eqs.(\ref{eq:z1}) 
and (\ref{eq:z2}) ($z_1 = z_2 =0$).
Identifying $n_A$ and $\beta_A$ with 
$n$ and $\beta$, respectively, this is nothing more than the information 
presented in the preceding section
for obtaining the error exponent optimized  with respect to the
distribution $\vec{Q}$. 
Therefore, our SM-based framework is consistent 
with the result for $\alpha_A^* (D,R)$
reported in the IT literature. 

\subsubsection{$\alpha_B(D,R)$}

We next consider $\alpha_B(D,R)$ for $R < \Rcpd$.
In the IT literature, the exponent for the best 
code for $R < \Rcpd$
is given \cite{Csiszar} as 
\begin{eqnarray}
\alpha_B^* (D,R) = \underset{\vec{U} }{ \mathrm{min}}~
     KL(\vec{U}||\vec{P}) + |R(\vec{U},D) -R |^+\ln 2, 
\label{eq:Csiszar}
\end{eqnarray}
where $|x|^+ = x$ for $x \ge 0$, and is 0 otherwise.
This can be separately expressed as
\begin{numcases}
{\mathrm{min} }
\underset{\vec{U} : R \ge R(\vec{U},D)}{ \mathrm{min}}
     KL(\vec{U}||\vec{P}),   \label{eq:suc1}\\
\underset{\vec{U} : R \le R(\vec{U},D)}{ \mathrm{min}}
     KL(\vec{U}||\vec{P}) + (R(\vec{U},D) -R) \ln2.  
\label{eq:suc2}
\end{numcases}

As well as eq.(\ref{eq:Marton}), eq.
(\ref{eq:suc1}) is converted to the minimization of
\begin{eqnarray}
J_{B1}(\vec{U}, \vec{V},z_1,z_2) &=&
\sumj U(j) \ln \left[ \frac{U(j)}{P(j)} \right] \cr
& & + n_{B1} \left[ \left\{
\sumj \suml V(l|j) U(j) \ln \left[ \frac{V(l|j)}{Q(l)}
\right]  -R \ln 2 - z_1 \right\} \right. \cr
& & + \beta_{B1} \left\{
\sumj \suml V(l|j) U(j) d(j,l) - D - z_2
\right\}  \cr
& &+ \left. \sumj  \nu (j) \left\{
\suml V(l|j) - 1
\right\}
\right]
+ \xi \left\{ \sumj U(j) - 1 \right\},  \label{eq:JB1}
\end{eqnarray}
with respect to $\vec{U}, \vec{V}, z_1$, and $z_2$.
Although constraints $R < \Rcpd$ and $z_1 \le
0$ are different from those for $\alpha_A^*(D,R)$, 
the minimum is also achieved on the boundary in this case, 
which indicates $n_{B1} > 0~(z_1 =0, \partial J_{B1} /
\partial z_1 = - n_{B1} <0)$.
This means that the distribution $\vec{U}$ and 
the conditional distribution $\vec{V}$ can be 
represented as 
\begin{eqnarray}
U(j) &=& 
\frac{
P(j) \left\{  \suml
Q(l)  e^{-\beta_{B1} d(j,l) }  \right\}^{n_{B1}}
}
{
\sumj
P(j) \left\{  \suml
Q(l)  e^{-\beta_{B1} d(j,l) }  \right\}^{n_{B1}}
},~~~~~\forall j \in Y, 
\label{eq:Inf_UB1} \\
V(l|j) &=& \frac{ Q(l) e^{- \beta_{B1} d(j,l)}}{ \suml Q(l) e^{- \beta_{B1} d(j,l)}
},~~~~~\forall j \in Y, l \in \t{Y}, \label{eq:Inf_VB1}
\end{eqnarray}
using the Lagrange multipliers $n_{B1}$ and $\beta_{B1}$.

Minimization (\ref{eq:suc2}) can be rewritten as
\begin{eqnarray}
\underset{\vec{U}, \vec{V}: \sumj \suml V(l|j) U(j) d(j,l) \le D,~I \ge R \ln 2}
{\mathrm{min}}~KL(\vec{U}|| \vec{P}) + I -R\ln 2,
\end{eqnarray}
where $I$ is the mutual information expressed in eq.(\ref{eq:I}).
This can also be solved by the method of Lagrange
multipliers, which yields the minimization of
\begin{eqnarray}
J_{B2}(\vec{U}, \vec{V},z_1,z_2) &=&
\sumj U(j) \ln \left[ \frac{U(j)}{P(j)} \right] \cr
& & +
\sumj \suml V(l|j) U(j) \ln \left[ \frac{V(l|j)}{Q(l)}
\right]  -R \ln 2  \cr
& & + n_{B2} \left\{
\sumj \suml V(l|j) U(j) \ln \left[ \frac{V(l|j)}{Q(l)}
\right]  -R \ln 2 - z_1 \right\} \cr
& & + \beta_{B2} \left\{
\sumj \suml V(l|j) U(j) d(j,l) - D - z_2
\right\}  \cr
& &+ \sumj  \nu (j) \left\{
\suml V(l|j) - 1
\right\}
+ \xi \left\{ \sumj U(j) - 1 \right\},  \label{eq:JB2}
\end{eqnarray}
where $n_{B2}, \beta_{B2}, \{ \nu(j) \}$ and $\xi$ are Lagrange
multipliers, and $z_1~(\ge 0)$ and $z_2~(\le 0)$ are defined in
eqs.(\ref{eq:z1}) and (\ref{eq:z2}).
Note that $n_{B2} \le 0$ and $\beta_{B2} > 0$, because
$KL(\vec{U}||\vec{P}) + I - R \ln 2$ is not a convex function and 
we cannot exclude the possibility that $n_{B2} = 0$.

If the minimization (\ref{eq:suc1}) 
(or (\ref{eq:JB1}) ) is achieved for $0<n_{B1} \le 1$, 
the minimization (\ref{eq:suc2}) ( or (\ref{eq:JB2}) ) 
is achieved by the same $\vec{U}$ and $\vec{V}$, 
by setting $n_{B2} = n_{B1} - 1~(\le 0)$. 
However, if $n_{B1} > 1$, no distributions 
that minimize (\ref{eq:suc1}) can simultaneously be the solution of 
eq.(\ref{eq:suc2}), 
which indicates that eq.(\ref{eq:suc2})
is achieved by a distribution $\vec{U}$ that satisfies 
$R < R(\vec{U},D)$. In this case, $n_{B2}$ must be zero, 
and therefore eq.(\ref{eq:JB2}) is reduced to 
\begin{eqnarray}
J_{B2}(\vec{U}, \vec{V},z_1,z_2) &=&
\sumj U(j) \ln \left[ \frac{U(j)}{P(j)} \right] \cr
& & +
\sumj \suml V(l|j) U(j) \ln \left[ \frac{V(l|j)}{Q(l)}
\right]  -R \ln 2  \cr
& & + \beta_{B2} \left\{
\sumj \suml V(l|j) U(j) d(j,l) - D - z_2
\right\}  \cr
& &+ \sumj  \nu (j) \left\{
\suml V(l|j) - 1
\right\}
+ \xi \left\{ \sumj U(j) - 1 \right\}.  \label{eq:JB22}
\end{eqnarray}
Differentiating eq.(\ref{eq:JB22}) 
with respect to $V(l|j)$ and
$U(j)$ yields
\begin{eqnarray}
V(l|j) &=& \frac{ Q(l) e^{- \beta_{B2} d(j,l)}}{ \suml Q(l) e^{- \beta_{B2} d(j,l)}
},~~~~~\forall j \in Y, l \in \t{Y}, \label{eq:Inf_VB2} \\
U(j) &=& 
\frac{
P(j) \suml
Q(l)  e^{-\beta_{B2} d(j,l) }
}
{
\sumj
P(j)  \suml
Q(l)  e^{-\beta_{B2} d(j,l) }  
},~~~~~\forall j \in Y.
\label{eq:Inf_UB2}
\end{eqnarray}

Based on the above argument, 
the optimal exponent $\alpha_B^* (D,R)$ is 
assessed by the following procedure.
First, we employ ABA for the solution of
minimization (\ref{eq:suc1}) with respect to 
$n_{B1}> 0$, $\beta_{B1} > 0$ and $\vec{Q}$ 
using eqs.(\ref{eq:Inf_UB1}), 
(\ref{eq:Inf_VB1}) and (\ref{eq:Inf_Q}) 
under the constraints (\ref{eq:z1}) and
(\ref{eq:z2})~($z_1 = z_2 =0$).
If the solved $n_{B1}$ satisfies 
$0 < n_{B1} \le 1$, 
it is guaranteed that the obtained $\vec{U}$ and $\vec{V}$ 
achieve the minimization (\ref{eq:Csiszar}).
However, if the obtained $n_{B1}$ is greater than 1, 
this solution is not appropriate, because
minimization (\ref{eq:suc2}) is not achieved. 
Therefore, we have to search for another solution 
using eqs.(\ref{eq:Inf_UB2}), 
(\ref{eq:Inf_VB2}) and (\ref{eq:Inf_Q})
with $\beta_{B2} > 0$, under the 
constraint (\ref{eq:z2})~($z_2 = 0$), 
which can also be performed by ABA. 
In this case, the other constraint 
(\ref{eq:z1})~($z_1 > 0$) is always satisfied for
$n_{B1} > 1$, which is confirmed by 
the fact that the minimization of 
\begin{eqnarray}
\underset{\vec{U}:R \ge R(\vec{U},D)}{\mathrm{min}}~
KL(\vec{U}||\vec{P}) + (R(\vec{U},D) - R) \ln 2
\end{eqnarray}
can be achieved by an internal point with 
respect to $z_1$ if and only if $0 < n_{B1} \le 1$.

This procedure is identical to that of 
the RM-based approach presented in a previous section. 
Therefore, the framework developed in this paper 
is consistent with the result 
for $\alpha_B^* (D,R)$ reported in the IT literature.

\subsection{Discussion}
Here, two points are worth noting.
First, we have shown that the exponents
assessed by the RM-based method 
become identical to those of the {\em best code} 
in the IT literature, when optimized with 
respect to the code ensemble. 
However, this may be somewhat curious because 
$\alpha_{\{A,B\}}(D,R)$ characterizes
either the {\em average} of the compression failure or the success 
probability over a code ensemble, which implies
that $\alpha_{\{A,B\}}(D,R)$ does not necessarily coincide with 
the exponent of the best code, even if the 
ensemble is optimized. In order to examine 
a possible difference in 
exponents between the average and optimal 
probabilities, we evaluated the exponents of 
the minimum failure probability 
$P^*_{\rm \tiny F}=\lim_{t \to -\infty}
\left \langle P^t_{\rm \tiny F}({\cal C},D) 
\right \rangle_{\mathcal{C}}^{1/t}$ 
for $R >\Rcpd$ and
the maximum success 
probability $P^*_{\rm \tiny S}=\lim_{t \to +\infty}
\left \langle P^t_{\rm \tiny S}({\cal C},D) \right 
\rangle_{\mathcal{C}}^{1/t}$ for $R <\Rcpd$
for fixed ensembles, which 
reduced to the current calculations for the average probabilities. 
This means that in the RCE specified by $\vec{Q}$,
the performance of the best code is identical to 
that of typical codes in terms of the exponents, 
although differences may exist for ensembles of other types. 
Second, we may be able to apply the present framework to 
sources with memory, for which the optimal exponents 
have not been reported in the IT literature. This possibility
is currently under investigation.

\section{Application to a Sub-optimal Ensemble}
In addition to consistency with the existing 
results, a major advantage of 
the proposed RM-based approach is its ability to 
accurately evaluate the exponents for a wider class of 
ensembles. Here, we demonstrate this ability for a lossy compression of 
a binary memoryless source, which is specified 
by $\vec{P}=(P(0),P(1))=(1-p,p)$ where $0 < p< 1/2$. 

Although RCEs exhibit the optimal performance, 
they are difficult to implement in practice because a storage 
of $O(M \! \times \! 2^N)$ is 
required in order to express 
the set of representative vectors 
$\t{\vec{y}}(\vec{s})$. 
As a candidate to resolve this difficulty, 
we investigate the performance
of a compression scheme which utilizes
perceptrons having random connections \cite{Hosaka}.

More specifically, we define a map from 
the compressed expression 
$\vec{s} \in \{+1,-1\}^N$ to 
the representative sequence 
$\tilde{\vec{y}}(\vec{s}) \in \{0,1\}^M$ as
\begin{eqnarray}
\tilde{y}^\mu(\vec{s})=f \left(\frac{1}{\sqrt{N}} 
\vec{s} \cdot \vec{x}^\mu \right),~~~~~(\mu=1,2,\ldots,M)
\label{eq:perceptron}
\end{eqnarray}
for the specification of a code, 
where $f(\cdot)$ is a function for which 
the output is limited to $\{0,1\}$
and $\vec{x}^{\mu=1,2,\ldots,M}$ are randomly 
predetermined $N$-dimensional vectors 
generated from an $N$-dimensional normal distribution 
$P(\vec{x}) = \left (\sqrt{2 \pi} \right )^{-N} \exp 
\left [-|\vec{x}|^2/2 \right ]$. 
These vectors are known to the compressor and 
the decompressor, which act as the 
codebook. Here, for convenience, we introduce the alphabet $\{+1,-1\}$, 
rather than the conventional 
alphabet $\{0,1\}$ with respect to the compressed sequence \vec{s}. 

We employ the Hamming distortion 
$d( \vec{y}, \t{\vec{y}}(\vec{s})) = \sum_{\mu=1}^M \left[ 
1 - \delta_{y^\mu, \t{y}^\mu} \right]$
to measure the fidelity of the representative sequences. 
Then, a lossy compression scheme can be defined on the
basis of eq.(\ref{eq:perceptron}) as follows: 
\begin{itemize}
\item {\bf Compression:} 
For a given message $\vec{y}$, find 
a vector $\vec{s}$ that minimizes the distortion 
$d(\vec{y},\tilde{\vec{y}}(\vec{s}))$, 
where $\tilde{\vec{y}}(\vec{s})$ is the representative vector
that is uniquely generated from $\vec{s}$ by eq.(\ref{eq:perceptron}). 
The obtained $\vec{s}$ is adopted as the compressed expression. 
\item {\bf Decoding:} 
Given the compressed expression $\vec{s}$, 
the representative vector $\tilde{\vec{y}}(\vec{s})$ produced by 
eq.(\ref{eq:perceptron}) yields an approximation of  
the original message. 
\end{itemize}
Random selection of the connections naturally 
defines a code ensemble of this scheme.

Codes of this type may be preferred for practical implementation
because the necessary storage cost is only $O(M \! \times \! N)$. 
However, possible correlations between components 
of the representative vector may prevent the analysis of its performance by 
conventional methods in the IT literature. Nevertheless, 
the proposed RM-based approach makes it possible 
to accurately evaluate the performance 
of this ensemble using a recipe similar 
to the capacity analysis of perceptrons, 
which has been reported extensively over the last decade 
\cite{perceptrons}. 
In a previous paper \cite{Hosaka}, 
such an analysis indicated that a function $f(u) = 1$
for $|u|<k$, and $0$ otherwise, which offers 
optimal performance in the limit $M,N \to \infty$
achieving the rate-distortion function of 
$R(p,D)=H_2(p)-H_2(D)$ for this case, 
where $H_2(x)=-x\log_2(x)-(1-x)\log_2(1-x)$
for $0<x<1$ 
when $k$ is adjusted such that 
$2 \int_{k}^\infty dz e^{-z^2/2}/\sqrt{2\pi} =
\frac{1-D^* - p}{1-2D^*}$,
where $D^*$ represents the lower bound of
the Hamming distortion for a given compression 
rate $R$, which is obtained from the
inverse function of the rate distortion function,
except for a very narrow AT instability region 
in the vicinity of $p=0.5$.

The error exponents of this ensemble 
can be calculated by a procedure similar to that for the RCE. 
Taking the average of $\left (
\sum_{\vec{s}}e^{-\beta d(\vec{y},\tilde{\vec{y}}(\vec{s}))}
\right )^n$ with respect to 
the original message $\vec{y}$ and 
the connection vectors $\vec{x}^{\mu=1,2,\ldots,M}$
yields
\begin{eqnarray}
&& g(n, \beta) =  \cr
&& \hspace{-0.8cm}
\mathop{\rm extr}_{ \{ q_{ab} \} }
\left \{
- \frac{1}{M}
\ln 
\left [
\int \left( \prod_{a=1}^n dv_a du_a \right)
\exp \left(
-\frac{1}{2} \vec{v}^t A \vec{v} + i \vec{v} \cdot \vec{u}
\right) 
\left \langle 
\prod_{a=1}^n  \!
e^{-\beta}\!+\!(1-e^{-\beta}) \Theta_k(u^a; y)
\!
\right \rangle_{y}^M
 \right. \right. \cr
&& \left. \left . 
\times 
\mathop{\rm Tr}_{\vec{s}^1,\! \vec{s}^2,\!\ldots, \! \vec{s}^n}
\prod_{a>b}
\delta \left ( \vec{s}^a \cdot \vec{s}^b - N q_{ab} \right )
\right ]   \right \}, 
\label{eq:gnperceptron}
\end{eqnarray}
where $q_{ab} =\frac{\vec{s}^a \cdot \vec{s}^b}{N}$
for $a>b=1,2,\ldots,n$, 
\begin{eqnarray}
      \Theta_k(u;1)&=& 1-\Theta_k(u;0) = \left\{
	\begin{array}{ll}
		1, & \mbox{for $|u| \le  k$} \\
		0, & \mbox{otherwise},
	\end{array}
	\right.  
\end{eqnarray}
and $A=\left (\delta_{ab}+(1-\delta_{ab})q_{ab} \right )$. 
This expression corresponds to eq.(\ref{eq:replica}) 
for the RCE.

The mirror symmetry, $f(-u) = f(u)$, 
of the transfer function yields a 
solution $q_{ab}=q=0$ under the RS 
ansatz \cite{Hosaka}, which offers 
\begin{eqnarray}
g_{\rm \tiny RS1}(n, \beta) 
=  
- \ln \left[ p \left\{ 1 - \eta + \eta e^{-\beta}   \right\}^n
+  (1-p) \left\{ (1-\eta) e^{-\beta} + \eta   \right\}^n \right] 
- n R \ln 2,  \label{eq:percep_sol0}
\end{eqnarray}
 where $\eta$ is defined as 
$\eta = 1 - 2 \int_{k}^\infty dz e^{-z^2/2}/\sqrt{2\pi}$.
In addition, there exists another RS solution
\begin{eqnarray}
g_{\rm \tiny RS2}(n, \beta)  
= 
- \ln \left[ p \left\{ 1-\eta + \eta e^{-\beta n}   \right\}
+  (1-p) \left\{ (1-\eta) e^{-\beta n} + \eta   \right\} \right] 
-R \ln 2,    \label{eq:percep_sol1}
\end{eqnarray}
corresponding to $q_{ab}=q=1$. 
Equations (\ref{eq:percep_sol0}) and 
(\ref{eq:percep_sol1}) coincide with 
eqs.(\ref{eq:sol0}) and (\ref{eq:sol1}) 
for the current source and the Hamming distortion, 
respectively.

\begin{figure}[t]
\begin{minipage}{0.45\linewidth}
\begin{center}
\includegraphics[width=6cm,height=5cm]{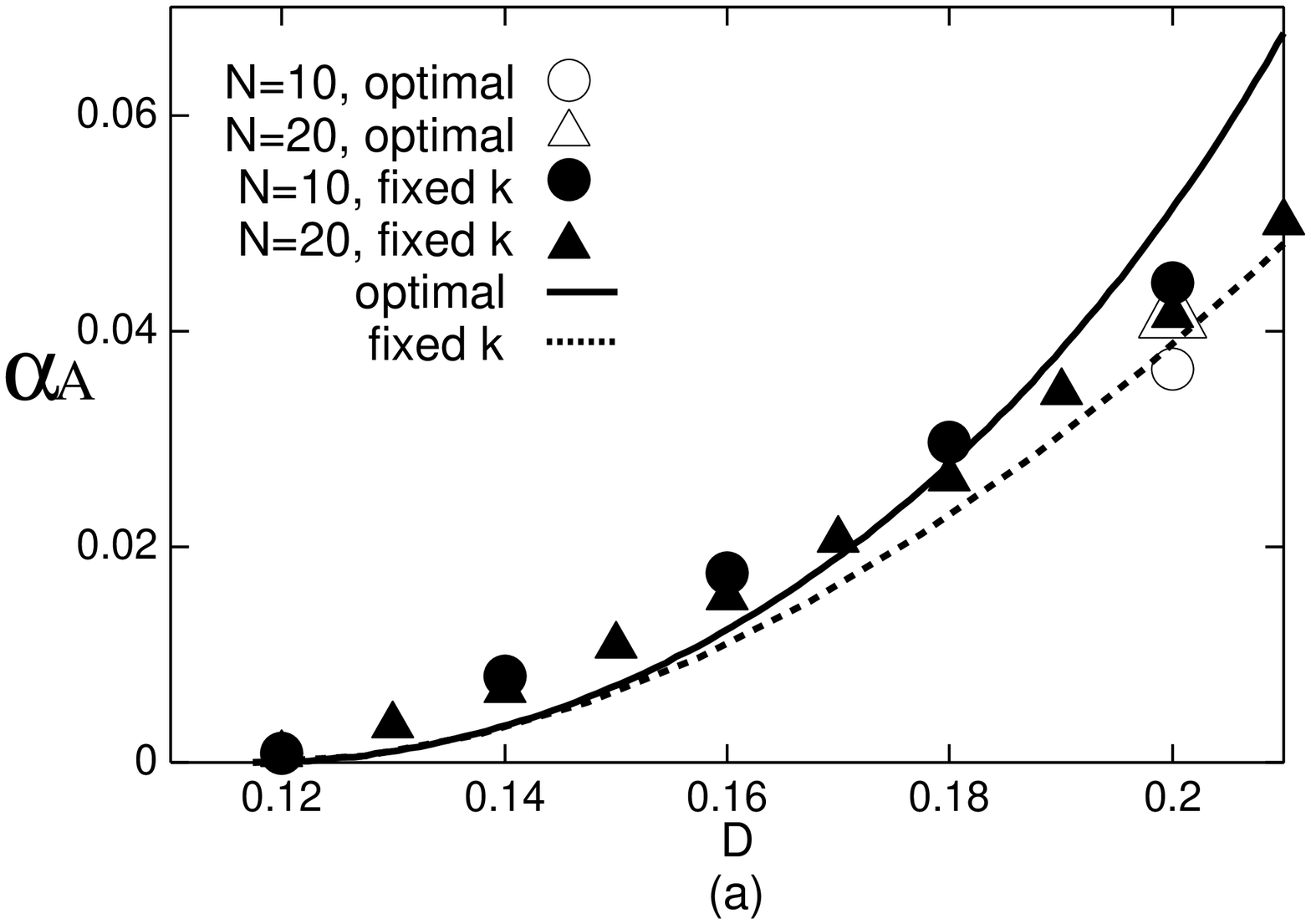}
\end{center}
\end{minipage}
\hspace{0.5cm}
\begin{minipage}{0.45\linewidth}
\begin{center}
\includegraphics[width=6cm,height=5cm]{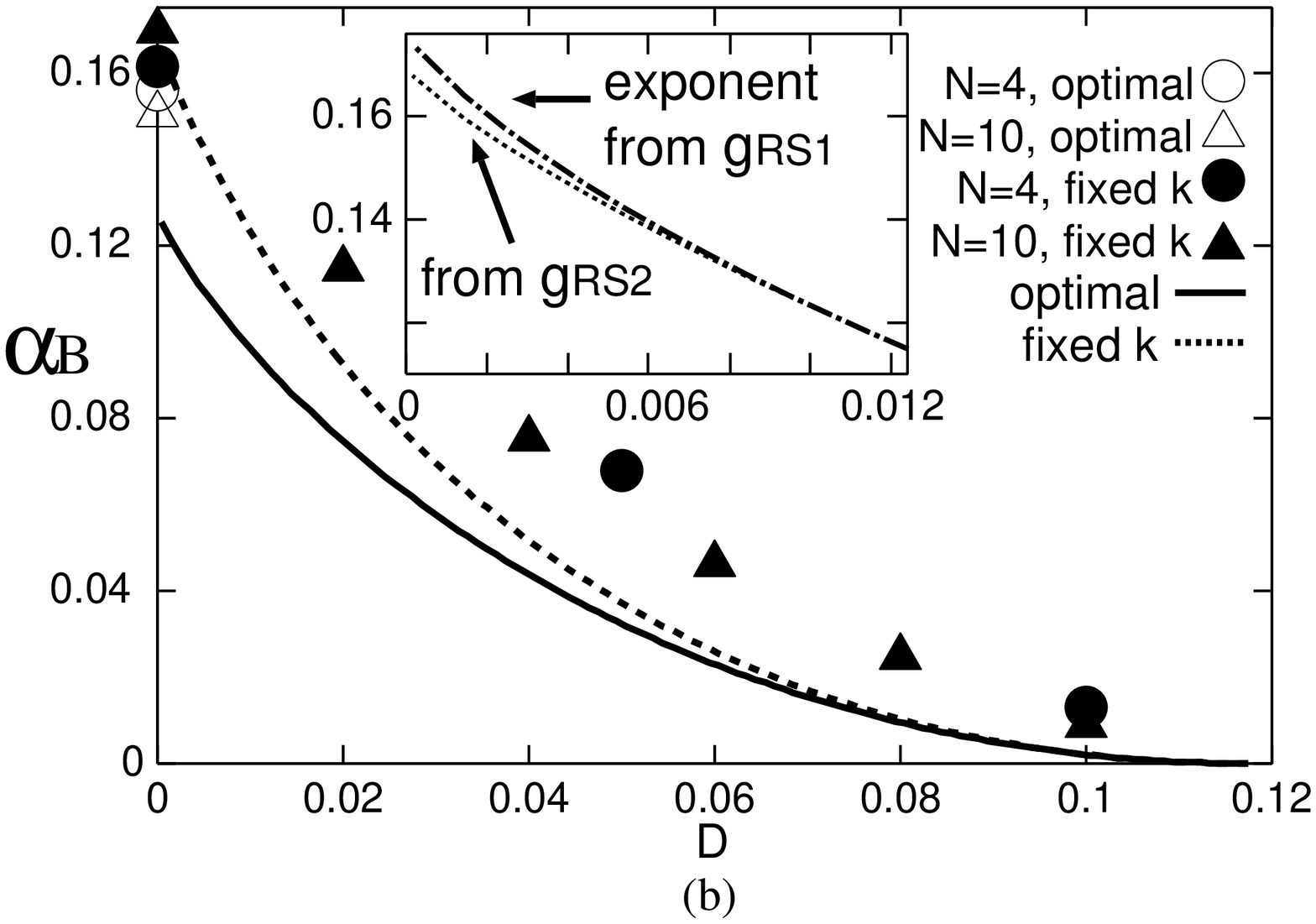}
\end{center}
\end{minipage}
\caption{Error exponents (a) $\alpha_A(D,R)$ and
(b) $\alpha_B(D,R)$ for $p=0.2,~R=0.2$.
The solid and dashed curves indicate  
the optimal exponents $\alpha_{\{A,B\}}^*(D,R)$ and 
the exponents obtained for fixed $k(\simeq 0.136)$ that realizes the 
rate-distortion relation 
$R(p,D)=H_2(p)-H_2(D)$, respectively. 
For $0 \le D \ltsim 0.011$ in Fig. (b), the dashed curve is obtained
from the solution $g_{ \mathrm{RS2} }(n, \beta)$,
which dominates $g_{ \mathrm{RS1} }(n, \beta)$ in this region (Fig. (b) inset).
The experimental data was obtained for 
(a) 5000-10000 trials for $N=10, 20$ 
and  
(b) $10^6$ trials for $N=4, 10$ 
through exhaustive search. 
The white circles and triangles represent the exponents optimized for (a) $D=0.2$ 
and (b) $D=0.0$, respectively, by adjusting $k$, 
and the black symbols indicate exponents for fixed $k(\simeq 0.136)$.
}
\label{fig:exp}
\end{figure}

Therefore, we can recycle the calculation for the RCE
to examine the performance of the current ensemble, 
which indicates that the optimal error exponents 
can be obtained by adjusting the parameter $k$ 
to the optimal value for each pair of $D$ and $R$ 
(such that $2 
\int_{k}^\infty dz e^{-z^2/2}/\sqrt{2\pi} =
\frac{1-D - p^*}{1-2D}$, where $p^*$ 
satisfies the relation $R = H_2(p^*) - H_2(D)$ 
for the rate $R$ and the given permissible level $D$) 
unless AT instability occurs for the 
above RS solutions. Note that for the current source, 
the optimal exponents can always be achieved by
$g_{\rm \tiny RS1}(n,\beta)$. Here,
$g_{\rm \tiny RS2}(n,\beta)$ becomes
dominant in only suboptimal 
cases for $\alpha_B(D, R)$ (Fig.~\ref{fig:exp} 
(b) inset).

In order to justify the above analysis, 
we performed numerical experiments 
implementing the proposed scheme. 
As an exhaustive search was performed for compression, 
the system size was limited to $N\!=\!20$. 
Fig.~\ref{fig:exp} shows the exponents 
averaged over the results from $5 
\times 10^3 \sim 1 \times 10^6$ experiments 
for the case of $p=0.2, R=0.2$. 
The white circles and triangles indicate data obtained by 
adjusting $k$ so that the exponents are optimized 
for (a) $D=0.2$ and (b) $D=0.0$, respectively. 
The black circles and triangles indicate data obtained using $k \simeq 0.136$ so as to reproduce the rate-distortion relation, which 
implies that both exponents vanish at $D^*\simeq 0.117$. 
In Fig.~\ref{fig:exp}(a), 
notice that the 
white symbols increase at $D=0.2$ as $N$ 
grows, whereas the black symbols decrease, 
approaching each theoretical prediction consistently. 
Fig.~\ref{fig:exp} (b) shows that the white symbols 
are located below the black symbols at $D=0.0$. 
In both figures, the discrepancies between the experimental data and the theoretical predictions are considered to be due to the finite size effect.

\section{Summary}
In summary, we have developed a scheme by which to assess
the error exponents of a lossy data compression problem
using RM. 
The newly developed RM-based approach 
for the exponents corresponding to the average failure or success
probabilities for the random code 
ensembles reproduces the optimal error exponents 
achieved by selecting the best
code reported in the IT literature, 
which indicates that the performance
of the best code is identical to that of typical codes 
in terms of error exponents. 
Furthermore, the proposed framework makes
an accurate assessment of the coding performance 
possible for a wide class of code ensembles. 
Using this characteristic, we have shown that a lossy 
compression scheme based on a specific 
type of non-monotonic perceptron 
provides the optimized exponents 
in most cases, which has been supported numerically.

Evaluation of 
the error exponents of practical
algorithms for lossy data compression is a subject for future study.
In order to reduce the computational cost of the proposed coding scheme, 
the development of approximation algorithms by which to realize the compression 
phase using a perceptron is currently under way.

\section*{Acknowledgment}
TH would like to thank T. Uematsu for his helpful comments. 
TH is a Research Fellow of the Japan Society for the Promotion of
Science.
The present study was supported in part by Grants-in-Aid (No.~164453 from the JSPS:TH) and (No.~14084206 from MEXT, Japan:YK).


\end{document}